\documentclass{kapproc} 

\normallatexbib

\begin{document}


%
%
%
\def\th{\theta}         \def\Th{\Theta}
\def\ga{\gamma}         \def\Ga{\Gamma}
\def\be{\beta}
\def\al{\alpha}
\def\ep{\epsilon}
\def\la{\lambda}        \def\La{\Lambda}
\def\de{\delta}         \def\De{\Delta}
\def\om{\omega}         \def\Om{\Omega}
\def\sig{\sigma}        \def\Sig{\Sigma}
\def\vphi{\varphi}
%
%
\def\CA{{\cal A}}       \def\CB{{\cal B}}       \def\CC{{\cal C}}
\def\CD{{\cal D}}       \def\CE{{\cal E}}       \def\CF{{\cal F}}
\def\CG{{\cal G}}       \def\CH{{\cal H}}       \def\CI{{\cal J}}
\def\CJ{{\cal J}}       \def\CK{{\cal K}}       \def\CL{{\cal L}}
\def\CM{{\cal M}}       \def\CN{{\cal N}}       \def\CO{{\cal O}}
\def\CP{{\cal P}}       \def\CQ{{\cal Q}}       \def\CR{{\cal R}}
\def\CS{{\cal S}}       \def\CT{{\cal T}}       \def\CU{{\cal U}}
\def\CV{{\cal V}}       \def\CW{{\cal W}}       \def\CX{{\cal X}}
\def\CY{{\cal Y}}       \def\CZ{{\cal Z}}
\def\debut{ \begin{eqnarray} }
\def\fin{ \end{eqnarray} }
\def\non{ \nonumber }
%



\articletitle[]{A Classification of Non-Hermitian Random Matrices~\footnote{Published in the Proceedings of the NATO Advanced Research Workshop on Statistical Field Theories, Como, Italy, 2001; ISBN 978-1-4020-0761-3; DOI https://doi.org/10.1007/978-94-010-0514-$2\_19$.}}


\chaptitlerunninghead{Non-Hermitian Random Matrices}

\author{Denis Bernard}
\affil{Service de physique th\'eorique, CE Saclay,
F-91191 Gif-sur-Yvette, France.}
\email{dbernard@spht.saclay.cea.fr}

\author{Andr\'e LeClair}
\affil{Newman Laboratory, Cornell University, Ithaca, NY 14853.}
\email{leclair@mail.lns.cornell.edu}




\begin{abstract}
We present a classification of non-hermitian random matrices based 
on implementing commuting discrete symmetries. It contains 38 classes.
This  generalizes the classification of hermitian random matrices due to
Altland-Zirnbauer and it also extends the Ginibre ensembles of non-hermitian
matrices. 
\end{abstract}




Random matrix theory originates from the work
of Wigner and Dyson on random hamiltonians \cite{Dyson}.
Since then it has been applied to a large variety of
problems ranging from enumerative topology, combinatorics,
to localization phenomena,  fluctuating surfaces, 
integrable or chaotic systems, etc...
Non-hermitian random matrices also have applications
to interesting quantum problems such as
open choatic scattering, dissipative quantum maps,
non-hermitian localization, etc... See e.g. ref.\cite{revue}
for an introduction. The aim of this short note
is to extend the Dyson \cite{Dyson} and Altland-Zirnbauer \cite{AltZirn}
classifications of hermitian random matrix ensembles to 
the non-hermitian ones~\footnote{Compared to our 2001 version, we corrected two misprints in one of the following tables that in the previous version led us to miscount the number of classes as 43 whereas it should have been 38.  Explicit details of the classification are unchanged.}.

\section{What are the rules?}
As usual, random matrix ensembles are constructed by
selecting  classes of matrices with specified 
properties under discrete symmetries \cite{Dyson,Mehta}. 
To define these ensembles we have to specify 
(i) what are the discrete  symmetries,
(ii) what are the equivalence relations among the matrices, 
and (iii) what are the probablility measures for each class.

(i) \underline{What are these discrete  symmetries.}\\
Let $h$ denote a complex matrix.
We demand that the transformations specifying 
random matrix classes are involutions ---
their actions are of order two. 
So we consider the following set of symmetries~\footnote{Any symmetry involving complex conjugation can be chosen as time reversal.}:
\debut
{\rm C\ sym.}&:&\quad h = \ep_c\, c\, h^T \, c^{-1},\quad
c^Tc^{-1}=\pm {\bf 1}
\label{Csym}\\
{\rm P\ sym.}&:&\quad h = - p\, h \, p^{-1},\quad 
p^2={\bf 1}
\label{Psym}\\
{\rm Q\ sym.}&:&\quad h =  q\, h^{\dag} \, q^{-1},\quad 
q^{\dag}q^{-1}={\bf 1}
\label{Qsym}\\
{\rm K\ {\rm sym.}}&:&\quad h = k\, h^* \, k^{-1},\quad 
kk^*=\pm {\bf 1}
\label{Ksym}
\fin
$h^T$ denotes the transposed matrix of $h$, $h^*$ its
complex conjugate and $h^{\dag}$ its hermitian conjugate.
The factor $\ep_c$ is just a sign $\ep_c=\pm$.
We could have introduced similar signs in the definitions of
type $Q$ and type $K$ symmetries;  however they can 
be  removed by redefining $h \to ih$.

We also demand that these transformations are 
implemented by {\it unitary} transformations:
\debut
cc^{\dag}={\bf 1},\ pp^{\dag}={\bf 1},\ 
qq^{\dag}={\bf 1},\ kk^{\dag}={\bf 1}
\fin

In the case of hermitian matrices one refers 
to type $C$ symmetries as particle/hole
symmetries or time reversal symmetries depending on whether  $\ep_c=-$
or $\ep_c=+$ respectively.
Matrices with type $P$ symmetry are said to be  chiral.
Both type $Q$ and type $K$ symmetries 
impose reality conditions on $h$ and they are  
redundant for
hermitian matrices.

\medskip

(ii) \underline{ What are the equivalence relations.}\\ 
We consider matrices up to {\it unitary} changes of basis, 
\debut
 h\to u\,h\,u^{\dag} \label{equiv}
\fin 
In other words, matrices linked by unitary similarity transformations
are said to be gauge equivalent.
For the symmetries (\ref{Csym}--\ref{Ksym}), this gauge equivalence
translates into:
\debut
c\to ucu^T,\ p\to upu^{-1},\ q\to uqu^{\dag},\ k\to uku^{*-1}
\label{symgauge}
\fin

The classification relies heavily on this rule and on
the assumed unitary implementations of the discrete symmetries.

We shall only classify minimal classes, which by definition 
are those whose matrices do not commute with a fixed matrix. 

\medskip

(iii) \underline{What are the probability measures.}\\
Since each of the classes we shall describe below is a subset of 
the space of complex matrices,
the simplest probability measure $\mu(dh)$ one may choose is 
obtained by restriction of the gaussian one defined by
\debut
 \mu(dh) = {\cal N}\, \exp( - {\rm Tr}hh^{\dag}\, )\, dh \label{gauss}
\fin
with $\CN$ a normalization factor.
It is invariant under the map (\ref{equiv}).

\medskip

There is  of course some degree of arbitrariness in formulating these
rules, in particular concerning the choice of the gauge equivalence (\ref{equiv}). 
It however originates on  one hand by requiring the gaussian
measure (\ref{gauss}) be invariant, and one the  another hand from
considering auxiliary hermitian matrices $\cal H$ obtained by
doubling the vector spaces on which the matrices  $h$ are acting.
These doubled matrices are defined by:
\debut
\CH = \pmatrix{ 0 & h\cr h^{\dag} & 0\cr}
\label{double}
\fin
They are always chiral as they anticommute with $\ga_5={\rm diag}(1,-1)$. 
Any similarity transformations $h\to uhu^{-1}$ are  mapped into
$\CH\to\CU\CH\CU^{\dag}$ with $\CU={\rm diag}(u,u^{\dag\,-1})$.
So, demanding that these transformations also act by similarity
on $\CH$ imposes $u$ to be unitary.

On $\CH$, both type $P$ and $Q$ symmetries act as chiral
transformations, $\CH\to -\CP\CH\CP^{-1}$ with $\CP={\rm diag}(P,P)$
and $\CH\to \CQ\CH\CQ^{-1}$ with $\CQ=\pmatrix{0&Q\cr Q&0\cr}$,
and $\CH$ may be block diagonalized if $h$ is $Q$ or $P$ symmetric.
Indeed, if $h$ is $Q$ symmetric then $\CQ$ and $\CH$
 may be simultaneously diagonalized since they commute. 
If $h$ is $P$ symmetric, $\CH$ commutes with the product
$\ga_5\CP$.

Type $C$ and $K$ symmetries both act as particle/hole
symmetries relating $\CH$ to its transpose $\CH^T$.
The classification of the doubled hamiltonians $\CH$ thus reduces
to that of chiral random matrices, cf. \cite{AltZirn}.
However, the spectra of $h$ and $\CH$ may differ significantly so that
we need a finer classification involving $h$ per se. 

\section{Intrinsic definition of classes.}
To specify classes we demand that the matrices belonging 
to a given class be invariant under
one or more of the  symmetries (\ref{Csym}--\ref{Ksym}).
It is important to bear in  mind that 
when imposing two or more symmetries it is the group generated
by these symmetries which is meaningful. Indeed, these groups 
may be presented in various ways depending on which
generators one picks. For instance, if a matrix possesses
both a type $P$ and a type $C$ symmetry, then it automatically
has another type $C$ symmetry with $c'=pc$ and $\ep'_c=-\ep_c$.

The intrinsic classification concerns the classification of the
symmetry groups generated by the transformations (\ref{Csym}--\ref{Ksym}).

We demand, as usual, that the transformations (\ref{Csym}--\ref{Ksym})
commute.
For any pair of symmetries the commutativity conditions read:
\debut
c=\pm pcp^T &;\quad p^*=\pm k^{-1}pk&;\quad 
q=\pm pqp^{\dag}\label{com1}\\
q^T=\pm c^{\dag}q^{-1}c &;\quad q^*=\pm k^{-1}qk^{{\dag}\,-1}
&;\quad k^Tc^{-1}kc^*=\pm {\bf 1} \non
\fin
The signs $\pm$ are arbitrary; they shall correspond to 
different groups.
\medskip

\underline{Without reality conditions.}

This arises if no type $K$ and type $Q$ symmetry
is imposed so that no reality condition is specified 
and $h$ is simply a complex matrix.
We may then impose either a type $P$ or a type $C$ symmetry or both. 
Not all groups generated by a type $P$ and a type $C$ symmetry
are distinct since, as mentioned above, the product of these
symmetries is another type $C$ symmetry but with an
opposite  sign $\ep_c$.
The list of inequivalent symmetry groups, together
with the inequivalent choices of the sign $\ep_c$, is the following:
\begin{center}
\begin{tabular}{ccc}
\hline
 Generators & Discrete symmetry group  & Number\\
    & Defining relations  & of classes \\
\hline\\
No sym  & no condition  & 1\\
$P$ sym & $p^2=1$ & 1  \\
$C$ sym & $c^T=\pm c,\ \ep_c$ & 4 \\
$P$, $C$ sym & $p^2=1,\ c^T=\pm c,\ pcp^T=c$ & 2\\
$P$, $C$ sym & $p^2=1,\ c^T=\pm \ep_c \, c,\ pcp^T=-c$ & 2\\
 ~& ~ \\
\hline
\end{tabular}
\end{center}
The 10 cases in this table correspond to the Altland-Zirnbauer classes.  
If the sign $\ep_c$ does not appear as an entry it
means that the value of this sign is irrelevent --- opposite
values correspond to identical groups. The sign factors
$\pm$ written explicitly are relevant --- meaning that 
e.g. the groups generated by a type $C$ symmetry with $c^T=c$ or
$c^T=-c$ are inequivalent.
The equivalences among the defining relations 
for groups generated a type $P$ and a type $C$ symmetry 
 are the following:
\begin{center}
\begin{tabular}{ccc}
$(p^2=1,c^T=\pm c, c = pcp^T)_{\ep_c}$ & $\cong$ &
$(p^2=1,c^T=\pm c, c = pcp^T)_{-\ep_c}$ \\
$(p^2=1,c^T=\pm \ep_c\, c, c =- pcp^T)_{\ep_c}$ & $\cong$ &
$(p^2=1,c^T=\pm\ep_c\, c, c =- pcp^T)_{-\ep_c}$ \\
\end{tabular}
\end{center}
In the above table we anticipate the numbers of
classes which we shall describe
in the following section. They depend on the numbers
of inequivalent representations of each set of defining relations.

Considering discrete groups generated by more symmetries of type $P$
or $C$ does not lead to new minimal classes. 
For instance, if the group is generated
by two type $P$ symmetries, then their product commutes with $h$ and
thus they do not define a minimal class. Similarly, suppose that we
impose two type $C$ symmetries with sign $\ep_{c1}$ and $\ep_{c2}$.
If $\ep_{c1}\ep_{c2}=-$, their product makes a type $P$ symmetry and
the group they generate is among the ones listed above. 
If $\ep_{c1}\ep_{c2}=+$, their product commutes with $h$ and
they do not specify a minimal class. 
More generally, considering more combinations of type $P$ and
type $C$ symmetries does not lead to new minimal classes as
in such cases one may always define fixed matrices commuting
with $h$.

\medskip

\underline{With reality conditions.}

This arises if we impose at least a type $Q$ or a type $K$ symmetry,
but we may simultaneously also impose symmetries of
other types. Again, there could be different but equivalent
presentations of the same discrete group as not all of these symmetries
are independent. For instance, a product of a type $P$ symmetry with
a type $C$, $Q$ or $K$ symmetry is again a type $C$, $Q$ or $K$.
The list of inequivalent groups generated by 
two or three of these symmetries
is the following:

\begin{center}
\begin{tabular}{ccc}
\hline
Generators & Discrete symmetry groups & Number \\
  & Defining relations & of classes \\
\hline\\
  
$Q$ sym        & $ q=q^{\dag} $ & 1 \\
$K$ sym     & $ kk^* = \pm  1 $ & 2 \\
$P$, $Q$ sym  & $ p^2=1,\ q^2=1,\ q=\pm pqp^{\dag}  $& 2 \\
$P$, $K$ sym & $ p^2=1,\ kk^* = \pm  1,\ kp^*=pk $& 2 \\
$P$, $K$ sym & $ p^2=1,\ kk^* = 1,\ kp^*=-pk $&  1 \\ 
$Q$, $C$ sym & $q=q^{\dag},\ c^T=\pm c,\ 
q^T=c^{\dag}q^{-1}c,\ \ep_c$& 4  \\
$Q$, $C$ sym & $q=q^{\dag},\ c^T=\pm c,\ q^T=-c^{\dag}q^{-1}c,\
\ep_c$& 4 \\
$P$, $Q$, $C$ sym & $p^2=1,\ q=q^{\dag},\ c^T=\pm c,\ \ep_c$& \\
~~ & $c=\ep_{cp}\, pcp^T,\ q=\ep_{pq}\, pqp^{\dag},\ 
     q^T=\ep_{cq}\,c^{\dag}q^{-1}c$& 12 \\
 ~ & ~ \\
\hline
\end{tabular}
\end{center}
The combined number of classes in the above two tables is $38$.
As above, the explicitly mentioned signs $\pm$ correspond 
to inequivalent groups.
The groups with defining relations
$( p^2=1,\ kk^* =1,\ kp^*=-pk)$ or 
$( p^2=1,\ kk^* =-1,\ kp^*=-pk)$ are equivalent. 
The groups generated either by a type $Q$ and a type $K$, or
by a type $C$ and a  type $K$ symmetries are included in this
list because the symmetries of type $C$, $Q$ or $K$ are linked
by the fact the product of two of them produces a symmetry of 
the third type.
The last cases, quoted in the last line of the above list,
are made of groups generated by three symmetries,
one of a type $P$ and two of type either $C$, $Q$ or $K$.
Their defining relations depend on the choices of the signs
$\ep_{cp}$, $\ep_{pq}$ and $\ep_{cq}$. The equivalences
between these choices are the following:
\begin{center}
\begin{tabular}{ccc}
$(c^T=\pm c; \ep_{cp}=\ep_{pq}=\ep_{cq}=+)_{\ep_c}$
& $\cong$ & $(c^T=\pm c; \ep_{cp}=\ep_{pq}=\ep_{cq}=+)_{-\ep_c}$ \\
$(c^T=\pm c; \ep_{cp}=\ep_{pq}=-\ep_{cq}=+)_{\ep_c}$
&$\cong$& $(c^T=\pm c; \ep_{cp}=\ep_{pq}=-\ep_{cq}=+)_{-\ep_c}$ \\
$(c^T=\pm c; \ep_{cp}=-\ep_{pq}=\ep_{cq}=+)_{\ep_c}$
&$\cong$& $(c^T=\pm c; -\ep_{cp}=\ep_{pq}=\ep_{cq}=-)_{-\ep_c}$ \\
$(c^T=\pm\ep_c\, c; -\ep_{cp}=\ep_{pq}=\ep_{cq}=+)_{\ep_c}$
&$\cong$& $(c^T=\pm\ep_c\, c; \ep_{cp}=-\ep_{pq}=\ep_{cq}=-)_{-\ep_c}$ \\
$(c^T=\pm\ep_c\, c; \ep_{cp}=\ep_{pq}=-\ep_{cq}=-)_{\ep_c}$
&$\cong$&$ (c^T=\pm\ep_c\, c; \ep_{cp}=\ep_{pq}=\ep_{cq}=-)_{-\ep_c}$ \\
\end{tabular}
\end{center}
Considering groups generated by more symmetries does not
lead to new minimal classes since in such cases one may construct
matrices commuting with $h$.
  
\section{Explicit realizations of the classes.}
Having determined the inequivalent groups of
commuting discrete symmetries, the second step consists
in finding all inequivalent representations of the defining relations 
of those groups. Due to the rules we choose, in particular the second
one, eq.(\ref{equiv}), we only consider representations in which
all symmetries are unitarily implemented and which
are not unitarily equivalent.
 
We shall list all these representations, adding an
index $\ep_c$ to recall when the action (\ref{Csym})
of the corresponding discrete group 
depends on $\ep_c$.
\medskip

\underline{Without reality conditions:}

If we impose no symmetry at all, the class is simply the set
of complex matrices. 

If we impose only a type $P$ symmetry. The matrix $p$ is unitary
and square to the identity, so that it is diagonalizable with
eigenvalues $\pm 1$. The solution $p=1$ is trivial as it implies
$h=0$. Assuming for simplicity that the numbers of
$+1$ and $-1$ eigenvalues are equal, 
we may choose a basis diagonalizing $p$:
\debut
( p = \sigma_z\otimes1); \label{p1}
\fin
Here and below, $\sigma_z,\ \sigma_x$ and $\sigma_y$ denote
the standard Pauli matrices.

If we only impose a type $C$ symmetry, $c$ is unitary and either
symmetric and antisymmetric. As is well known \cite{Dyson}, 
up to an appropriate gauge choice 
it may be presented into one of the following forms:
\debut
( c = 1)_{\ep_c};\quad (c=i\sigma_y\otimes 1)_{\ep_c}; \label{c1}
\fin

Let us now impose a type $P$ and a type $C$ symmetry.
In the basis diagonalizing $p$ with $p = \sigma_z\otimes1$, 
the commutativity relation $pcp^T=\pm c$ 
means that either $p$ and $c$ commute or anticommute.
If they commute, then $c$ is block diagonal in this basis so that
$c=1\otimes 1$ or $c=1\otimes i\sigma_y$ depending whether it
is symmetric or antisymmetric. If they anticommute, $c$ is
block off-diagonal so that, modulo unitary change of basis,
it may be reduced to $c=\sigma_x\otimes 1$ or $c=i\sigma_y\otimes 1$. 
However, as explained in the previous intrinsic
classification, these two cases correspond the same
symmetry group but with opposite signs $\ep_c$.
So a set of inequivalent representations is:
\debut  
&& (p=\sigma_z\otimes 1,\ c=1\otimes 1);
 (p=\sigma_z\otimes 1\otimes 1,\ c=1\otimes i\sigma_y\otimes 1)  ;\non\\
&&  ~~~~~~~~~~~~~~~~~~~
(p=\sigma_z\otimes 1,\ c=\sigma_x\otimes 1)_{\ep_c} ; \label{pc1}
\fin
Altogether there are $10$ classes of non-hermitian 
random matrices without reality conditions.
They are of course parallel to the $10$ classes of hermitian
random matrices \cite{AltZirn}.
\medskip

\underline{With reality conditions:}

Imposing only a type $Q$ symmetry, we have $q=q^{\dag}$ and
$qq^{\dag}=1$ since, by choice, we implement the symmetry
unitarily. So $q$ is diagonalizable with eigenvalues $\pm 1$.
The solution $q=1$ is non-trivial as it simply means that 
$h$ is hermitian. When $q$ possesses both $+1$ and $-1$ 
eigenvalues we assume that they are in equal numbers, hence:
\debut
(q=1);\quad (q=\sigma_z\otimes 1) ; \label{q1}
\fin

Imposing only a type $K$ symmetry, we have $kk^*=\pm1$,
$kk^{\dag}=1$, and their classification is similar to
that of type $C$ symmetries. There are two cases:
\debut
(k=1);\quad (k=i\sigma_y\otimes 1); \label{k1}
\fin

When imposing both  type $P$ and $Q$ symmetries, $p$ and $q$ 
both square to the identity and either commute and anticommute.
In the gauge in which $p=\sigma_z\otimes1$ the possible 
$q$ are $1\otimes1$, $\sigma_z\otimes1$ or $\sigma_x\otimes 1$,
up to unitary similarity transformations preserving the form of $p$.
However, the two first possibilities generate identical groups,
thus the inequivalent representations are:
\debut
 (p=\sigma_z\otimes 1,\ q=1\otimes 1);\quad
(p=\sigma_z\otimes 1,\ q=\sigma_x\otimes 1) ;
\fin

Let us next impose a type $P$ and a type $K$ symmetry.
With the unitarity property of $p$ and $k$, the commutativity
relations between these symmetries are similar to those
between type $P$ and $C$ symmetries. Thus, the classification 
of these representations may be borrowed from eq.(\ref{pc1})
and we have:
\debut 
&& (p=\sigma_z\otimes 1,\ k=1\otimes 1);
 (p=\sigma_z\otimes 1\otimes 1,\ k=1\otimes i\sigma_y\otimes 1)  ;\non\\
&&  ~~~~~~~~~~~~~~~~~~~
(p=\sigma_z\otimes 1,\ k=\sigma_x\otimes 1) ; \label{pk1}
\fin

When imposing  simultaneously a type $Q$ and a type $C$ symmetry,
their product is a symmetry of type $K$. The commutativity
constraints between type $Q$ and $C$  symmetries are solved in the same
way as the commutativity conditions for type $P$ and $C$ symmetries;
the only difference being that $q=1$ is now a non-trivial solution,
contrary to $p=1$. The inequivalent representations for $q$ and $c$ are:
\debut
(q=1,\ c=1)_{\ep_c} &;& 
(q=1\otimes1,\ c=i\sigma_y\otimes1)_{\ep_c} ; \label{qc1}\\ 
(q=\sigma_z\otimes1,\ c=1\otimes1)_{\ep_c} &;&
(q=\sigma_z\otimes1,\ c=1\otimes i\sigma_y)_{\ep_c}\non\\
(q =\sigma_z\otimes 1,\ c=i\sigma_y\otimes 1)_{\ep_c} &;&
(q=\sigma_z\otimes 1,\ c=\sigma_x\otimes 1)_{\ep_c};\non
\fin

Finally, let us impose 
a type $P$ symmetry together with two among 
the three types  of symmetries $Q$, $C$ and $K$.
The commutativity constraints are solved by first choosing
a gauge in which $p=\sigma_z\otimes1$.  The list of
inequivalent solutions is:
\debut
(p=\sigma_z\otimes1,\ & q =1\otimes1,\ 
&c=1\otimes1\ {\rm or}\ c=1\otimes i\sigma_y);\non\\
(p=\sigma_z\otimes1,\ & q =1\otimes1,\ 
&c=\sigma_x\otimes 1)_{\ep_c} ;\label{pcq1}\\
(p=\sigma_z\otimes1,\ & q =\sigma_x\otimes1,\ 
&c=1\otimes 1\ {\rm or}\ c=1\otimes i\sigma_y\ \non\\
& ~ & {\rm or}\ c=\sigma_x\otimes 1\ {\rm or}\ 
c=\sigma_x\otimes i\sigma_y)_{\ep_c};\non
\fin
Here, each choice of $c$ corresponds to a different class.

Altogether, eqs.(\ref{p1}--\ref{pcq1}) give $38$ classes 
of random non-hermitian matrices.
As for the Ginibre ensembles \cite{Gini} --- which correspond to the
classes without symmetry or with type $K$ symmetry with
$k=1$ or $k=i\sigma_y\otimes 1$ --- we expect that in the large
matrix size limit the density of states in each class covers a bounded
domain of the complex plane with the topology of a disc.





%



\begin{chapthebibliography}{<widest bib entry>}


\bibitem{Dyson} F. Dyson, J. Math. Phys. {\bf 3} (1962) 140;

\bibitem{Mehta} M. Mehta, {\it Random matrices}, Academic Press,
  Boston, 1991.

\bibitem{AltZirn} A. Altland and M. Zirnbauer, Phys. Rev. {\bf B 55}
  (1997) 1142; M. Zirnbauer, J. Math. Phys. {\bf 37} (1996) 4986;

\bibitem{Gini} J. Ginibre, J. Math. Phys. {\bf 6} (1965) 440;

\bibitem{revue} Y.V. Fyodorov and H.J. Sommers, J. Math. Phys. {\bf 38}
    (1997) 1918, and references therein;


\bibitem{Dirac} D. Bernard and A. LeClair, ``A classification of
2d random Dirac fermions'', J. Phys. {\bf A35} (2002) 2555-2567, [arXiv:cond-mat/0109552].

\end{chapthebibliography}
\end{document}